# Design of the readout electronics for the DAMPE Silicon Tracker detector

Fei Zhang(张飞)[1]   Wen-Xi Peng(彭文溪)[1]   Ke Gong(龚轲)[1]   Di Wu(吴帝)[1]   Yi-Fan Dong(董亦凡)[1]
Rui Qiao(乔锐)[1]   Rui-Rui Fan(樊瑞睿)[1]   Jin-Zhou Wang(汪锦州)[1]   Huan-Yu Wang(王焕玉)[1]
Xin Wu[2]   Daniel La Marra[2]   Philipp Azzarello[2]   Valentina Gallo[2]   Giovanni Ambrosi[3]   Andrea Nardinocchi[3;4]

1 Institute of High Energy Physics, Chinese Academy of Sciences, Beijing 100049, China
2 Département de Physique Nucléaire et Corpusculaire, University of Geneva, Geneva 1211, Switzerland
3 Istituto Nazionale di Fisica Nucleare Sezione di Perugia, Perugia 06100, Italy
4 Università di Perugia, Perugia 06100, Italy

**Abstract**: The Silicon Tracker (STK) is a detector of the DAMPE satellite to measure the incidence direction of high energy cosmic ray. It consists of 6 X-Y double layers of silicon micro-strip detectors with 73,728 readout channels. It's a great challenge to readout the channels and process the huge volume of data in the critical space environment. 1152 Application Specific Integrated Circuits (ASIC) and 384 ADCs are adopted to readout the detector channels. The 192 Tracker Front-end Hybrid (TFH) modules and 8 identical Tracker Readout Board (TRB) modules are designed to control and digitalize the front signals. In this paper, the design of the readout electronics for STK and its performance will be presented in detail.

## 1. Introduction

The Dark Matter Particle Explorer (DAMPE), which has been launched into the 500Km orbit on 17 December 2015, is a space science mission of the Chinese Academy of Science. Its main scientific objective is to detect 5Gev-10TeV electrons and photons in order to identify possible signatures of Dark Matter (DM). It will also measure the flux of nuclei up to 500TeV, which will bring new insights to the origin and propagation of the high energy cosmic rays. As illustrated in Fig. 1, from top to bottom, the DAMPE payload consists of four sub-detectors: a Plastic Scintillation Detector (PSD), a Silicon TracKer detector (STK), a BGO Calorimeter (BGO) and a Neutron Detector [1].

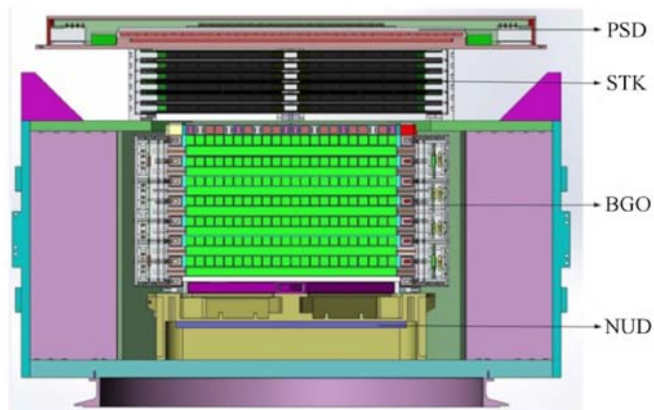

Fig. 1.   Architecture of the DAMPE payload

The STK detector was developed by an international collaboration composed of groups from Institute of High Energy Physics (China), University of Geneva (Switzerland), and INFN (Italy). It was designed to measure the charge of the nuclei cosmic rays, the charged particle tracks, as well as the photon direction. In order to achieve these goals, the 6 double X-Y orthogonal layers of high spatial resolution silicon micro-strip detectors and the internal Tungsten plates to convert incoming photons into electron/positron pairs were adopt in STK. As shown in Fig. 2, each tracking layer is made of 16 ladders each formed of 4 single-sided AC-coupled silicon micro-strip detectors (320μm thick and 121μ m pitch) for a total of 192 ladders [2]. There are 384 readout channels per ladder and for a total of 73728 channels for full STK. Due to the long term reliability in space, strict power supply and limited bandwidth of the satellite, it is a great challenge to readout all of the STK channels and process the huge volume of data on-board.

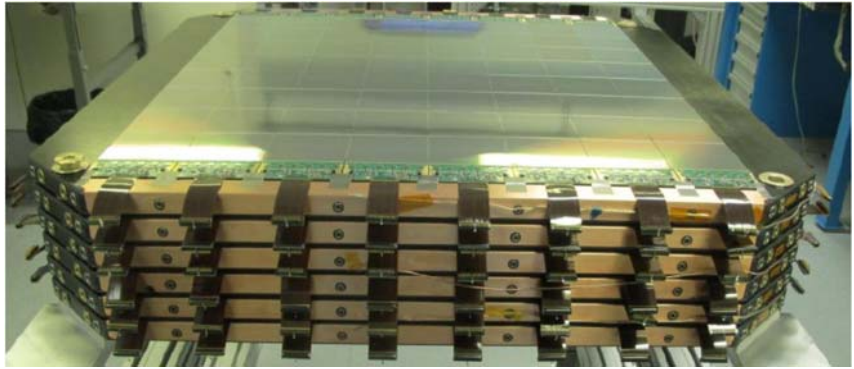

Fig. 2. Picture of STK detector layers before TRBs mounted

## 2. Requirements for readout electronics of STK

According to physics simulation, the peak charges of the Minimum Ionizing Particles(MIPs) in 300 μm silicon detector is 3.5fC. The RMS noise of the readout electronics is required to 0.3fC per silicon strip channel so that the STK can get a good resolution for the cosmic MIPs. In order to maximize the detecting efficiency on-board, the dead time after every event trigger is required to be less than 3 milliseconds.

More than 600GB raw data without data compression would be generated per day by STK at the mean 50 Hz event trigger rate of the in-orbit DAMPE. The volume of the STK scientific data output is limited to less than 8GB per day because of the limited downlink capability. Therefore on-board data compression technique is strongly required.

Restrictive power budget for STK is also put forward to the readout electronics because of the limited resource of the satellite. The STK power consumption is limited to less than 90W. In order to reduce the power consumption, many industrial grade electronic components with low power dissipation, high integration level and good performance were chosen for the readout electronics.

## 3. Readout electronics design
### 3.1 Overview of the readout electronics

The readout electronics for STK sub-detector consists of 192 Tracker Front-end Hybrid (TFH) modules and 8 identical Tracker Readout Board (TRB) modules. As shown in Fig. 3, every TRB connects 24 TFHs with a total of 9216 silicon strip channels. The public Payload Data Handling Unit (PDHU) receives all the hit signals of a random cosmic particle from the BGO sub-detector and generates a global trigger signal for all sub-detectors of the DAMPE [3]. Eight TRBs are powered by the public DC/DC modules and working in parallel under the control of the public PDHU. The housekeeping and scientific data of the TRBs are also transferred to the PDHU through the RS422 and LVDS datalinks, respectively.

To be published in Chinese Physics C

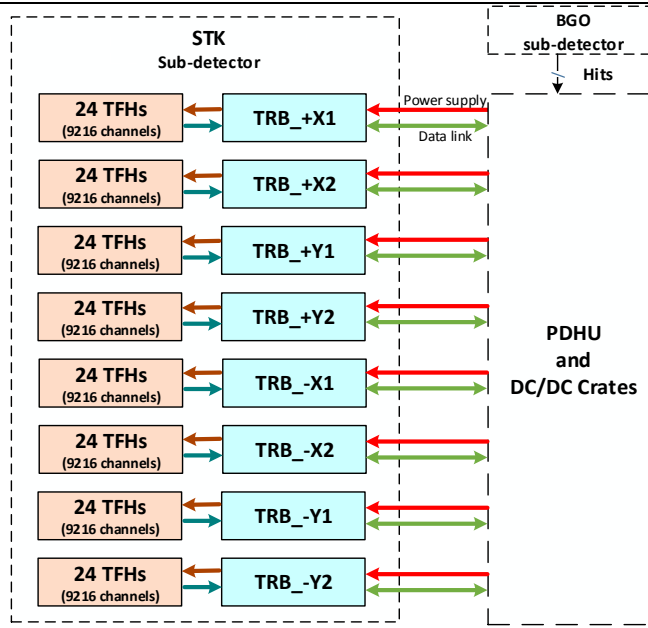

Fig. 3.    Architecture of the readout electronics for STK

The +X1 location TRB module is placed as illustrated in Fig. 4. The 24 TFHs in the 6 X-view layers are connected to the TRB_+X1 by their respective flexible hybrid cables. Other TRBs are also mounted like this in other 7 locations around the STK detector array.

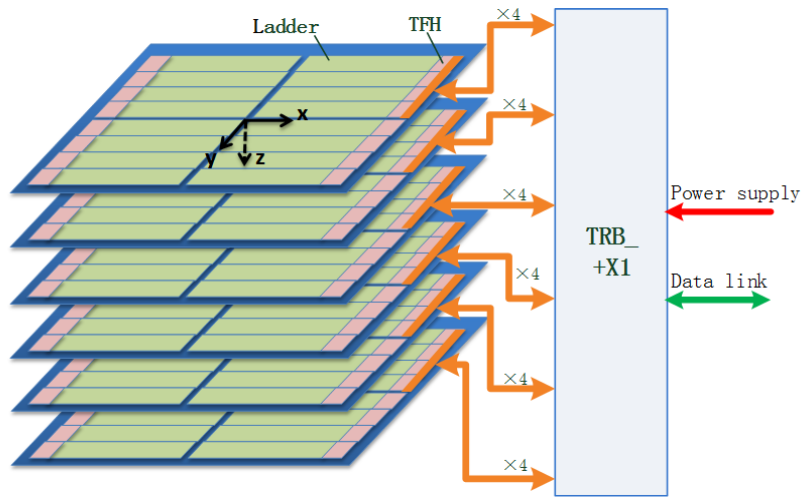

Fig. 4.    X-view detector arrangement and connection for +X1 TRB

The following Fig. 5 is a picture about the side view of the STK after two TRBs mounted in the framework of +X side.

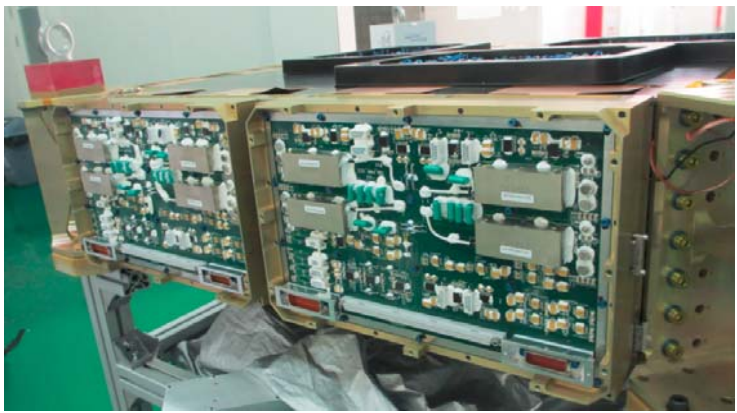

Fig. 5.    Side View of the +X TRBs mounted on the framework of STK

**3.1 TFH design**

The TFH is a kind of hybrid electronics board where a silicon strip ladder and its front readout electronics placed. The Application Specific Integrated Circuit (ASIC) VA140 was chosen to measure the charge when a cosmic particle hit the silicon strips. VA140 is a 64-channel, low noise, low power consumption (0.29mW/channel) and high dynamic range (+/-200fC) charge sensitive preamplifier-shaper ASIC designed by IDEAS Inc. (Norway) [4]. Our previous work about the prototype for cosmic-ray charge measurement based on VA140 connecting to the Si-PIN detectors was presented in the reference paper [5]. For DAMPE STK, 384 readout channels in every ladder are read out by six VA140 chips. The Chip-On-Board (COB) process and wire bonding technique were adopted to mount the ladder and bare VA140 chips to the TFH boards.

As illustrated in Fig. 6, the 6 VA140 chips in every TFH were divided into two groups. The three VA140 chips in every group cascaded and shared the driver signals and output amplifier circuits. Two mutual backup digital thermometers DS18S20Z were also adopted in every TFH to monitor the temperature. The power source VSS/VDD for the readout electronics and HV bias (80V) for silicon strip ladder in the TFH were supplied by its connected TRB.

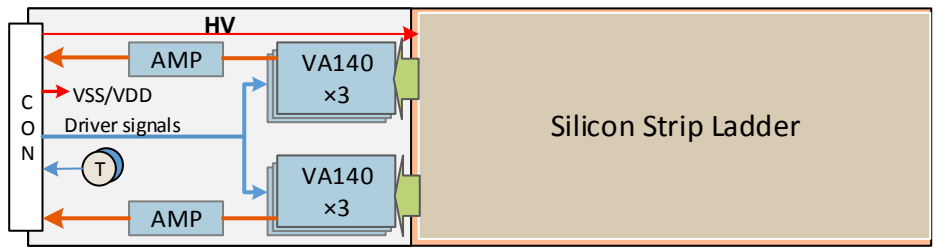

Fig. 6.    Block diagram of a TFH

**3.2  TRB design**

Eight identical TRB modules were mounted around the silicon detector array. They are responsible for the detector readout and data process for their front connected TFHs. Each TRB module consists of 3 electronics boards: power board, control board and SADC board. The block diagram of the TRB module is illustrated in Fig. 7, two FPGAs (called Master FPGA and slave FPGA, respectively) work together to control the data acquisition and communication. A SRAM (M65609E) is used to buffer the scientific data of every trigger and an EEPROM (EE1M08VS1192) is used to store the threshold parameters of each channel for data compression.

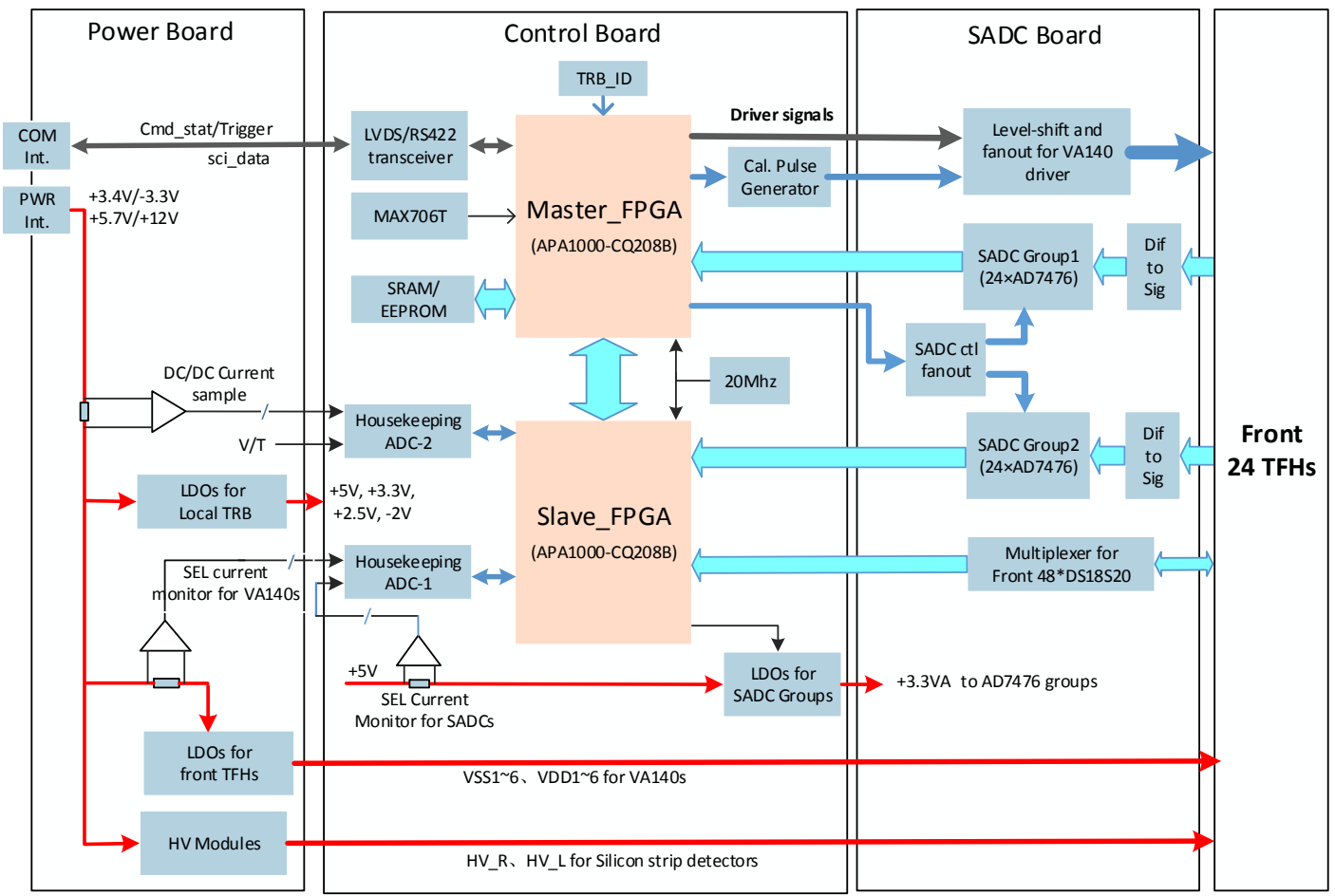

Fig. 7.  Block diagram of TRB module

The major parts of the TRB module will be described in detail in the following sections.

### 3.2.1 Power supply

The TRB modules are powered by the public DC/DC crates which converts the satellite primary power +28V to secondary power +3.4V, -3.3V, +5.7V and +12V. Due to the good noise ripple rejection of Low Dropout regulator (LDO), the adjustable positive and negative LDOs MSK5101 and LM2991S in the TRB power board were adopted to generate different voltages for front TFHs and local TRB electronic components.

Two HV-generator groups were also designed in the TRB power board. Each group supplied the HV bias for the front 12 detector ladders. The HV modules S9100P, which designed by SITAEL company, were adopted to generate the +80V HV biases. As shown in Fig. 8, two S9100P modules inside a HV-generator group are mutual backup and can be enabled or disabled by the FPGA. In order to protect the silicon strip detectors, a special circuit including a 200K ohm resistor, a 34uF capacitor and an amplifier LM6142 were designed before the Vset pin of S9100P to make the rising time of the HV voltage to be slower than 20 seconds.

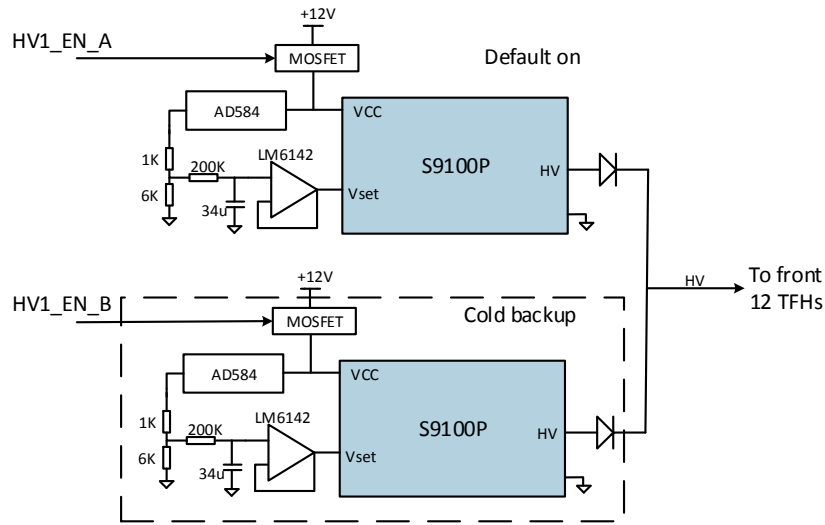

Fig. 8.   Circuit for a HV-generator group

### 3.2.2  Communication with PDHU

There are three kinds of communication mechanism between TRBs and public PDHU, as following description. Each of them has a respective redundant backup bus.

(1) Trigger bus. It is based on RS422 signal level and is falling edge effective. The PDHU will send the global trigger to TRBs through it when a coincidence event is determined during taking cosmic ray data.

(2) Command/state bus. It is based on RS422 level and Universal Asynchronous Receiver/Transmitter (UART) with 115200 baud rate and half duplex protocol. The PDHU sends remote commands and polls the house-keeping data to/from TRBs through it.

(3) Scientific data bus. It is based on LVDS level and user-defined serial protocol with 20MHz reference clock. The PDHU can accept maximum 2000 bytes scientific data from each TRB after every trigger.

### 3.2.3  VA140 driver and gain calibration

The VA140 chips in the front TFHs are read out simultaneously under the driver signals (CKB, HOLDB, DRESET, SHIFT_IN, TEST_ON) from TRB FPGAs. But the level of FPGA driver signals is 0 ~ 3.3V, while the level of VA140 is -2V ~ +1.5V. As illustrated in Fig. 9, the RS422 receiver chips (DS26LV32) were adopted to shift the level of the driver signals from FPGA level to VA140 level. The SN54LVTH162245 chips powered by -2V and +1.5V were also used to fan-out the driver signals so that the driver signals can be sent to the front TFHs separately. Even if one group of the driver signals for a TFH was failed, the others would not be affected.

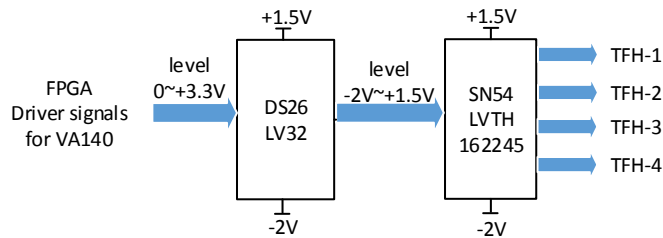

Fig. 9.   Level shift and fan-out circuit for VA140

A gain calibration circuit was designed to test the linearity of the VA140 channels by inject step pulses with different amplitudes to VA140 Calibration pad, as illustrated in Fig. 10. When the analog switch ADG201 is turned on, a step pulse will be generated by the circuit. Inside each VA140, a 2pF capacitor can convert the pulse to charge which is injected to every channel. The calibration channel can be selected by an analog de-multiplexer controlled by the bit-register of VA140. In order to increase the signal-to-noise ratio (SNR), the amplitude set by the DAC TLV5638 is 10 times larger than the value expected by the VA140, and the resistors 9K and 1K in the TFH board are used to divide the pulse amplitude. The inject charge can be calculated by this formula: $Q=0.1*V_m*2pF$, where $V_m$ is the amplitude set by the DAC.

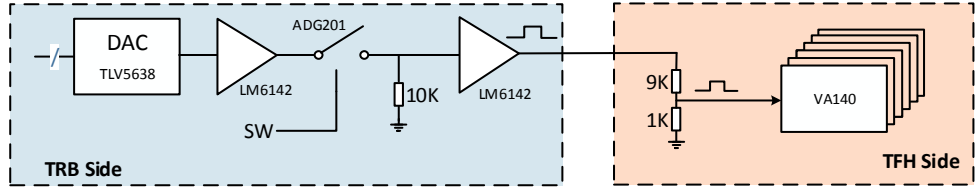

Fig. 10. Block diagram of Gain calibration circuit

### 3.2.4 Scientific data acquisition

In order to reduce the dead time of the VA140 digitization and make good use of the parallel work feature of FPGA, the STK readout task was divided into 384 parallel work sub-parts. As illustrated in Fig. 11, every sub-part consisted of three VA140 chips (connected to front 192 strip channels), two amplifiers (AD8032) and a Serial Analog-to-Digital Converter (AD7476AR). In the TFH board, every three VA140 chips cascaded under the control of a chain shifter register so that the 192 channels can be read out through the analog multiplexer successively. The output differential current drivers of the VA140 chips in one sub-parts are wired together and share the analog conditioning circuit. The analog conditioning circuit are used to buffer the differential current signals and convert them to single-end signal which can be digitized by a Serial ADC.

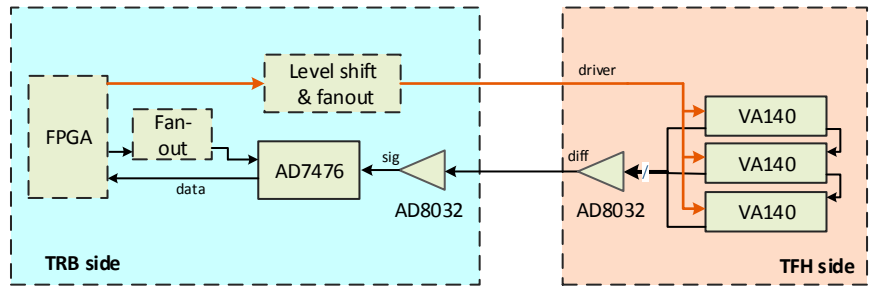

Fig. 11. Block diagram of a sub-part for VA140 digitization

### 3.2.5 House-keeping data acquisition

There are two kinds of house-keeping data collected by the TRBs and transferred to the PDHU through the command/state RS422 bus. One is the digital register values and statistical information in the FPGAs, the other is the analog information include currents, voltages and temperatures to monitor the status of the TRB and its connected TFHs. The analog house-keeping data include following items:

(1) The currents of each TRB power supply (+3.4V, -3.3V, +5.7V, +12V) and the SEL protecting groups (such as SADC group, VA140 group). They are sampled every seconds by the series sample resistors and amplifier LM6142s.
(2) The voltages of every group HV bias. They are sampled every 16 seconds by the divider resistors and amplifier LM6142.
(3) The temperatures. There are 48 digital thermometers (DS18S20Z) in the connected TFHs and 4 analog NTC thermistors (MF501) in the TRB boards.

### 3.2.6 FPGA software

In each TRB, two APA1000-CQ208B chips were adopted to control the data acquisition and communication. The FPGAs are Flash-bashed, non-volatile and MIL-STD-883B grade devices with features of high reliability and insensitivity to Single Event Latchup (SEL).

The logic block diagram of the TRB FPGAs is illustrated in Fig. 12. The identical Data_Process module in each FPGA is used to readout the 24 SADCs and process the scientific data for front 12 TFHs. The master FPGA is also responsible for sending driver signals to VA140, buffering the scientific data to the SRAM, storing/loading the threshold parameters to/from the EEPROM and communicating with the PDHU through RS422/LVDS bus. The slave FPGA is also responsible for the controlling of house-keeping data acquisition, HV supply and SEL protection.

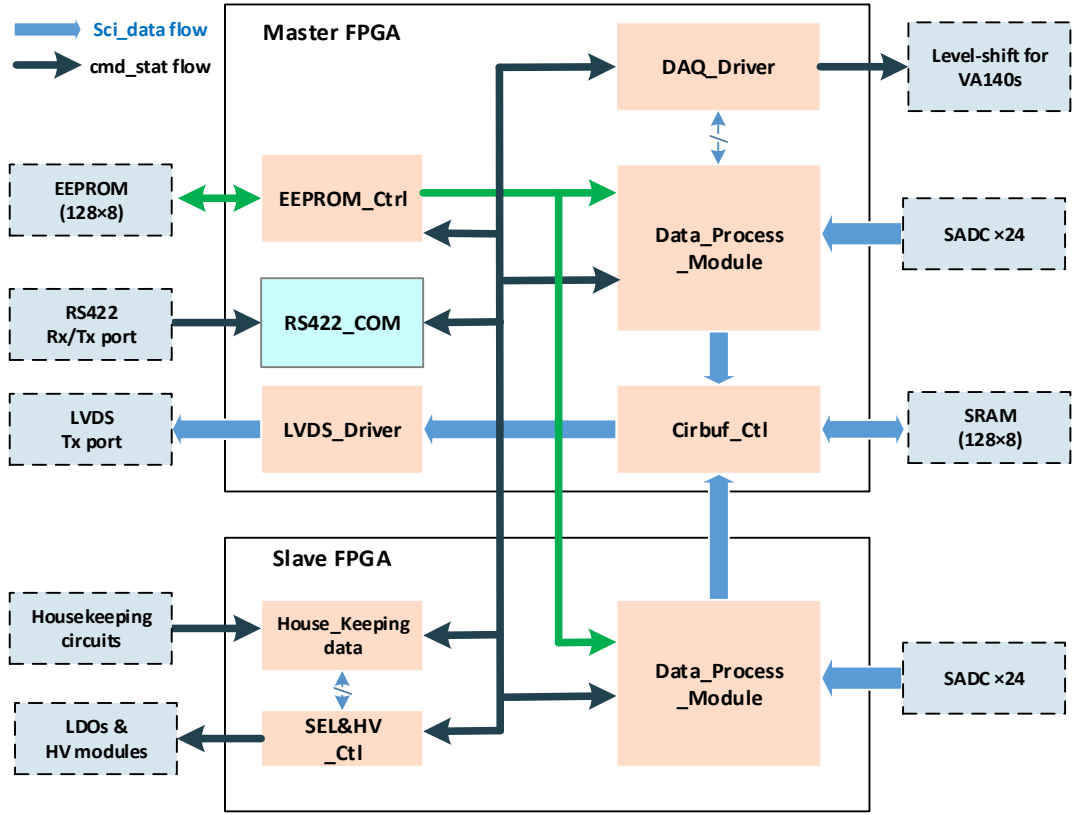

Fig. 12. Logic block diagram in TRB FPGAs

Four working modes of the scientific data acquisition are designed in the FPGAs, as descripted in table 1. The working mode can be selected by the PDHU commands through the RS422 bus.

Table 1. Description of STK working mode

| Working mode | description |
| --- | --- |
| raw data mode | Digitize the VA140 channels and transfer the raw data to the PDHU without any data compression |
| gain calibration mode | Test the linearity response of each VA140 channel by injecting different amplitudes of gain calibration pulses |
| pedestal update mode | Digitize the VA140 channels, calculate the average values of the 1024 times accumulation for each channel, and update the pedestal values stored in FPGAs |
| data compression mode | Digitize the VA140 channels, compress the data and then transfer to the PDHU |

Most of onboard time, the STK are working on data compression mode to detect the cosmic ray. According to the event trigger from PDHU, all the TRBs are working simultaneously to read out the STK detector signals, compress the data and transfer to the PDHU. The time distribution of the FPGA software for every event trigger is illustrated in Fig. 13. Benefiting from the rich logic resource and block ram of FPGAs, parallel and pipeline structure were used to process the STK data so that the dead time of every event trigger can be reduced to less than 2.95 milliseconds. The detailed on-board data compression algorithm realized in FPGA, including pre-processing (Pedestal subtraction, Common noise subtraction and bad channel cutting) and cluster finding, was presented in the reference paper [6].

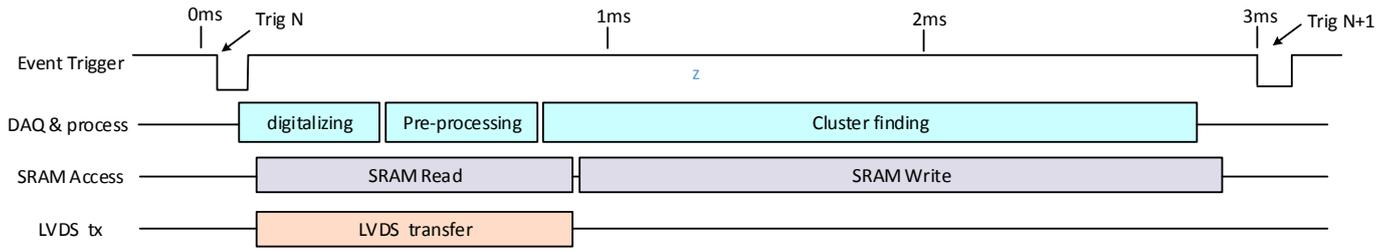

Fig. 13.   Time distribution for every event trigger

### 3.2.7   Mitigation of radiation effect

The critical space radiation environment causes radiation effects to the electronics, such as Total Ionizing Dose (TID), Single Event Latchup (SEL) and Single Event Upset (SEU). The anti-radiation design was taken into consideration both in the hardware and software design of the STK readout electronics.

The high grade electronic components with high TID and SEL tolerance were chosen for most parts of the STK readout electronics. But the industrial grade chips (VA140, AD7476AR and DS18S20Z), which are not immune to SEL effects [7], were also adopted because of their high integration, low power consumption and good performance.

In order to avoid the SEL damage, a SEL protecting method was designed for VA140 and AD7476AR chips. Every group of 24 VA140 chips in 4 TFHs, or every group of 24 AD7476AR chips in the TRB, share the power supply LDOs, respectively. Their power supply current is sampled every seconds by the Slave FPGA house-keeping logic. Once the current is larger than the configurable threshold, the Slave FPGA will control the LDOs to power off one second and power on again to eliminate the SEL. Another SEL protecting method was also designed for the DS18S20Z chips. A series resistor (100 ohm) was inserted in the power supply of every DS18S20Z chip to limit the current so that the DS18S20Z will be immune to the SEL [8].

Although the Actel Flash-based FPGA (APA1000) is SEU immune in the configuration logic elements because of its floating gate structure, but the SEU is still unavoidable in the D flip-flop registers and block RAMs inside FPGAs [9]. The Triple Modular Redundancy (TMR), Cyclic Redundancy Check (CRC), odd and cumulative checksum were adopted in the TRB FPGA software to check the upset bit and mitigate the harm of SEU.

## 4.   Performance test

### 4.1   Power consumption

For each TRB and its 24 connecting ladders, the current values of the power supply for +3.4V, -3.3V, +5.7V and +12V, were around 1600mA, 900mA, 350mA and 20mA, respectively. The STK consumes on average 83W of power during taking data (in data compression mode) at a nominal 50Hz trigger rate.

### 4.2   Gain Calibration Test

During STK working in Gain Calibration mode, ten interval input charges from 20fC to 200fC covering VA140 dynamic range were injected to the channels of the front VA140s. The calibration charge sweep response of the 384 channels in the first ladder connecting to TRB +X1 was shown in Fig. 14. According to the result of gain calibration test, the linearity parameter of the channels can be calculated. As illustrated in the linearity curve of the first channel, the INL is 3.21% and the gain is 0.063fC per ADC bin.

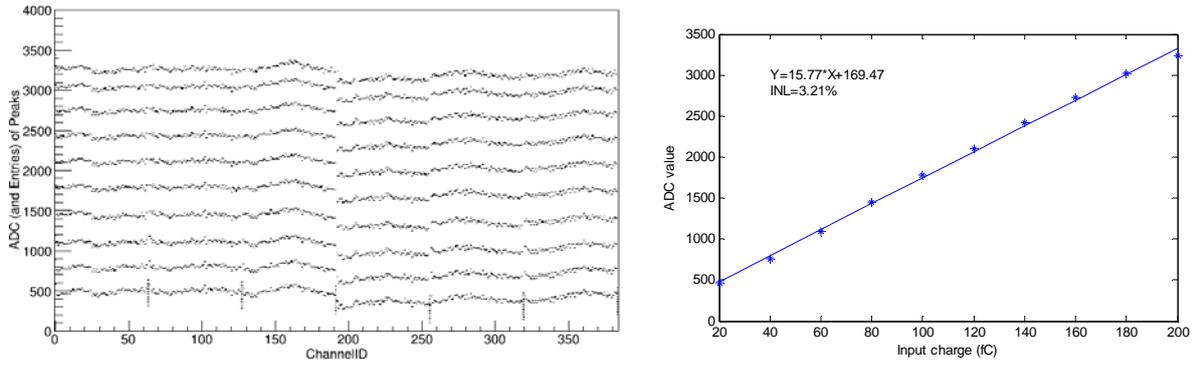

Fig. 14.   Gain Calibration overview of one ladder (Location: 1st ladder of TRB +X1 ) and linearity curve of the first channel

### 4.3 Pedestal and Noise test

During STK working in raw data mode, the 100Hz periodic trigger was used to test the pedestal and noise of the readout electronics. The pedestal distribution and RMS noise (after Common Noise subtracted) overview of full STK 73,728 channels was shown in Fig. 15. All the pedestal values are in the range 0 to 450 ADC bins and they obey the normal distribution. The RMS noise of most channels (around 99.6%) are less than 5 ADC bins. The noisy channels will be masked in the FPGA by setting a high threshold parameter for data compression.

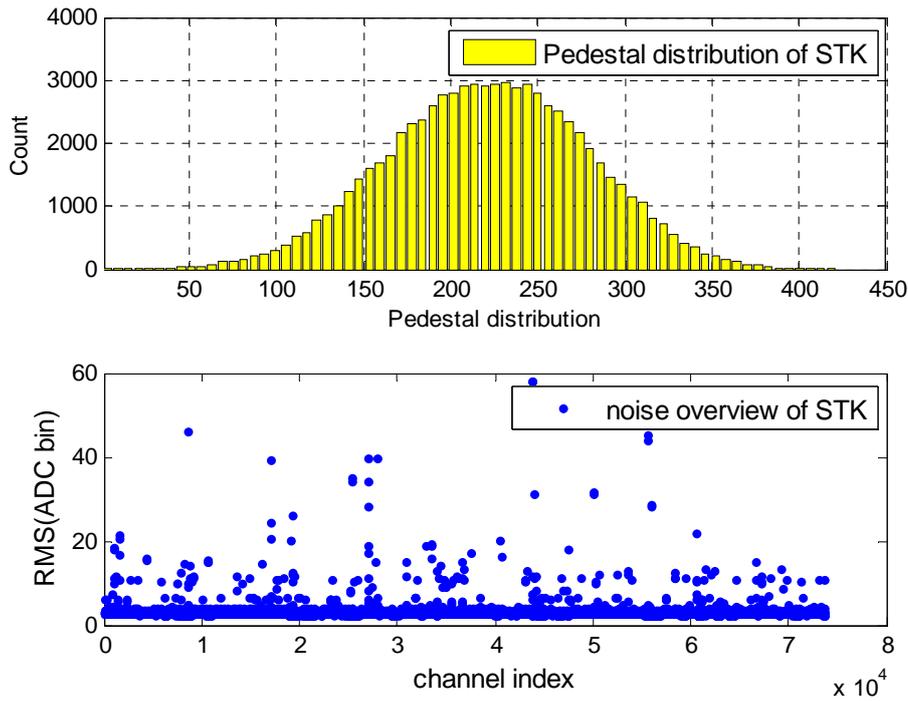

Fig. 15.   Pedestal distribution and noise overview of full STK 73728 channels

### 4.4 Cosmic ray test

During STK working in data compression mode, the coincidence event triggers were generated from the PDHU according to the BGO sub-detector hits of the ground cosmic rays (Muons). The one hour cumulative MIP curve of the cosmic rays collected by a TFH (ladder-02) were shown in Fig. 16. The MIP spectrum obeys a Landau distribution and can be clearly distinguished from the pedestal.

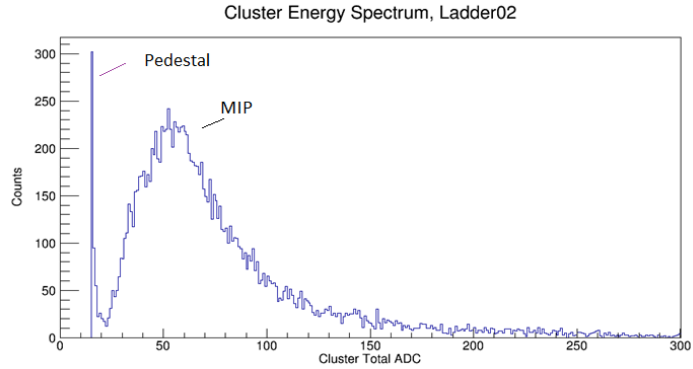

Fig. 16.　MIP curve of cosmic rays

　　The STK data packet length distribution of different cosmic rays was shown in Fig. 17. The mean length of event data packet is around 700 bytes, which is much smaller than the 150K bytes raw packets. The data compression rate is better than 0.5% for ground cosmic rays.

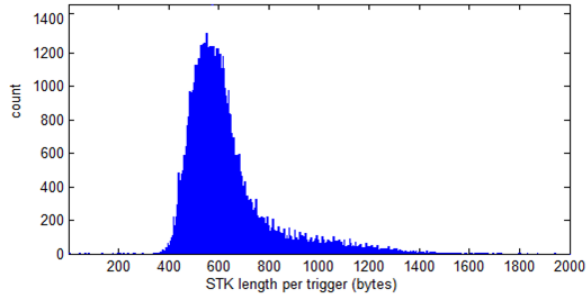

Fig. 17.　Event data packet length distribution of STK per trigger

## 5. Conclusion

　　Based on 8 TRB modules and 192 TFH modules, the readout electronics was successfully developed for DAMPE silicon tracker detector. As shown in Table 2, the performance of the readout electronics was demonstrated to meet all the design goals.

Table 2.　Readout electronics performance of STK

| Item | Performance |
| --- | --- |
| Power consumption | ~ 83W |
| RMS Noise | < 5 ADC bin (~0.3fC), 99.6% of all the channels |
| Integral Non-Linearity | ~ 3% during 200fC dynamic range of VA140 |
| Dead time | < 3ms |
| Data compression rate | better than 0.5% for ground cosmic rays |

　　After all the readout electronics modules assembled with the silicon strip detectors and fully tested (including the thermal-vacuum test, vibration test, EMC test et al.), the STK has been mounted to the satellite with other DAMPE sub-detectors on May 2015 and launched on December 2015. Currently the STK is keeping on taking data and its performance is quite stable as expected.